# Automated DC Voltage and DC Resistance Real-time Multiple Standard for Artifact Calibration of Calibrators and multimeters

[a]F. Galliana[a], P. Capra,[a] R. Cerri[a], C. Francese[a], M. Lanzillotti[a] L. Roncaglione Tet[a], F. Pennecchi[a] and A. Pollarolo[b]

[a] *Istituto Nazionale di Ricerca Metrologica (INRIM),*
*Str. delle Cacce 91–10135 Turin – Italy*

[b] *Measurements International Ltd. (MI), 118 Commerce Drive Prescott, Ontario, Canada.*
*E-mail:* `f.galliana@inrim.it`

ABSTRACT: An automated temperature-controlled electrical DC voltage and DC resistance multiple reference standard (MRS) has been developed by Measurements International (MI) with the scientific support from the Istituto Nazionale di Ricerca Metrologica (INRIM). The MRS includes a 10 V, a 1 Ω, and a 10 kΩ standards selectable via a switch unit. This setup allows the artifact calibration of high-end calibrators and multimeters used in low-frequency electrical measurements. The two resistors are high-stability standards from MI, while the 10 V standard is based on a low-noise circuit developed by INRIM in collaboration with MI. A key innovation is the internal real-time clock calendar, which displays the calibration values of the MRS standards and their updated values internally calculated. This ensures reliable use of the MRS standards over extended periods between calibrations, effectively minimizing uncertainties due to their drift. The standards are housed in a thermal box, minimizing temperature variations. The MRS standards meet the uncertainty requirements defined by calibrators and multimeters manufacturers for artifact calibration and can also serve as laboratory references or travelling standards for interlaboratory comparisons (ILCs). MI is currently commercializing the MRS.

**Contents**

1. **Introduction**

Standards and instruments require calibration with traceability to the International System of Units (SI) through national standards maintained by National Metrology Institutes (NMIs). Modern electrical digital instruments achieve traceability to the SI using a wide set of primary standards, such as DC voltage and resistors, DC and AC voltage dividers, AC/DC transfer standards, and current shunts, to cover wide measurement ranges. Alternatively, high-precision digital multimeters (DMMs), multifunction transfer standards (MTSs), or a limited set of primary standards can be employed [1]. In fact, high-end DMMs and multifunction calibrators (MFCs) can be calibrated with artifact calibration [2, 3], an in-house method requiring only three reference standards: 10 V, 1 Ω, and 10 kΩ. This procedure updates the internal references of DMMs and MFCs, which then automatically update other ranges and functions. However, since artifact calibration consists only of an adjustment of the references of the instruments, a full verification using a complete set of primary standards is periodically required to ensure their compliance with the manufacturer's specifications. At the Istituto Nazionale di Ricerca Metrologica (INRIM), portable temperature-controlled DC resistance setups for calibration of multifunction instruments were already realised [4, 5]. A first prototype of multiple DC Voltage and DC Resistance reference was developed at INRIM [6]. It was made up of two resistor networks and of a 10 V standard based on the LTZ 1000 circuit with the addition of 3D printed parts to maintain the stability of the integrated circuit at 48°C. After the development of this prototype, Measurements International (MI), with scientific and technical support from INRIM, has developed an automated, temperature-controlled, DC voltage and DC resistance real-time multiple reference standard (MRS) specifically designed for artifact calibration. The MRS includes a 10 V and a 1 Ω and a 10 kΩ resistors at single fruit. It employs a low-thermal-output scanner to select the artifact, minimizing electromotive forces (EMFs) and contact resistances during switching operations.

The scanner also allows the automatic polarity reversal of the 10 V reference. The MRS provides key advantages to the user, including effective thermal control and the ability to extrapolate real-time standard values. These features allow artifact calibrations to be performed at any time with uncertainties of the MRS standards comparable to their calibration ones. The MRS standards can also function as laboratory references in NMIs or travelling standards for interlaboratory comparisons (ILCs) [7], including high-level ones such as those in [8, 9].

## 2. The MRS structure

Figure 1 shows the block diagram of the MRS. The standards are housed in a temperature-controlled aluminium box, with the temperature monitored by an internal 100 Ω Platinum Resistance Thermometer (PRT). The MRS has an internal battery and supports remote control via RS-232 or GPIB interfaces. According to the manufacturer, a recommended stabilisation period of 12 hours is required after powering on the MRS to ensure that its standards are ready for accurate measurements [10]. The selected standard is connected to the output via a LEMO-brand push-pull connector and a scanner, either for its calibration or as a reference for artifact calibration.

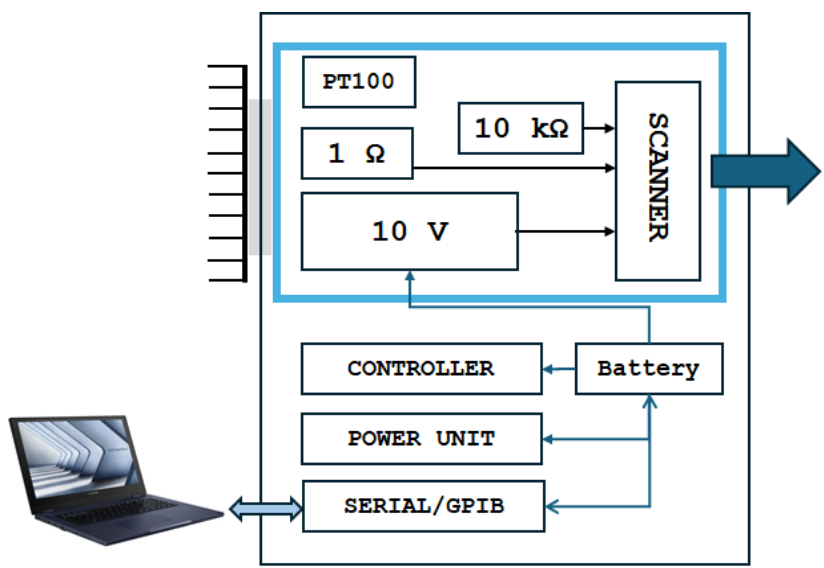

**Figure 1.** Simplified block diagram of the MRS. The three standards and the scanner are kept at constant temperature inside an electrostatic shield (blue box in the figure), and are separated from the electronic control and interfacing system and the power supply.

## 3. The MRS standards

Table 1 summarises the main specifications of the MRS standards, as provided in the manufacturer's data sheet.

Table 1. Features of the MRS Standards declared by the manufacturer.

| Standard | Temperature coefficient (TCR)[1] | Stability/Year | Dissipation |
| --- | --- | --- | --- |

---

[1] For ambient temperatures in the range (23 ± 5) C.

| | | | |
|---|---|---|---|
| 10 V | 0.1 μV/V/°C | 1.5 μV/V | |
| 1 Ω | 0.1 μΩ/Ω/°C | 0.5 μΩ/Ω | 10 mW |
| 10 kΩ | 0.1 μΩ/Ω/°C | 0.5 μΩ/Ω | 10 mW |

The resistors inside the MRS are manufactured by MI [10] and are based on well-established technologies. The 1 Ω resistor is a bifilar non-inductive resistive element with low and negligible temperature and pressure coefficients. Its voltage terminals are welded using Evanohm [11] filler metal to minimise EMF. An annealing procedure is employed to reduce the temperature coefficient (TCR) of the resistor and for structural stabilisation the element is sealed in an oil-filled copper case. The 10 kΩ resistor is instead of the metal foil type [12, 13]. The first prototype of the 10 V standard was developed at INRIM. It utilised a temperature-stabilised Zener diode (Linear Technology LTZ1000) maintained at 48 °C using an on-chip heater. Since the DC voltage reference is sensitive to thermal effects caused by contact points between the Zener pins and the printed circuit board tracks, the layout of the 10 V standard was specifically designed to minimise these effects. Within the MRS, the standards are anchored to a thermal mass maintained at 35 °C. This stable temperature minimises thermal drift, thereby enhancing the long-term stability of the standards. At INRIM, the TCR of the 10 V standard was evaluated between 18 °C and 28 °C by placing a MRS in a thermo-regulated air bath with a stability better than ±0.01 °C. Table 2 shows the temperature dependence of the three standards. The TCR of the resistors was verified in the same controlled air bath, but prior to their integration into the MRS to verify if their TCR was suitable for the requirements of the MRS project.

Table 2. Temperature dependence of three reference standards. All the measurements were made inside the INRIM climatic chamber. The 10 V was measured inside the MRS case, while the resistors were measured outside the MRS case.

| Temperature | Values of relative deviation from nominal value (×10⁻⁶) | | |
|---|---|---|---|
| (°C) | 10 V | 1 Ω | 10 kΩ |
| 18 | 1.91 | | |
| 19 | | | |
| 20 | 1.84 | | |
| 21 | | 9.8 | -9.49 |
| 22 | 1.85 | | |
| 23 | 2.04 | 10.1 | -9.51 |
| 24 | 1.93 | | |
| 25 | | 10.2 | -9.6 |
| 26 | 1.82 | | |
| 27 | | | |
| 28 | 1.84 | | |

The TCR of the 10 V standard has been conservatively estimated as approximately 22 nV/V/°C. The TCRs of the 1 Ω and 10 kΩ resistors, assuming linear drift, have been estimated as about 0.11 μΩ/Ω/°C and 0.16 μΩ/Ω/°C, respectively. Placing the resistors in the MRS box further improves their TCR, thanks to the thermal control. Table 3 shows the mean measurements of the 10 V standard of an MRS available at INRIM, maintained at 23 °C inside a thermal chamber over a period of about three months. The drift was estimated to be approximately 2.8 nV/V/day (assuming a linear drift), a result reasonably consistent with the annual stability specified by the manufacturer. More accurate predictions of the 10 V standard's annual stability could be achieved using the methods outlined in [14–17]. However, this has been considered to be outside the scope of the work since users of the MRS can update the stability of their standards at each calibration. Table 4 shows the drift of the 10 V standard without thermal control in a laboratory thermally controlled at (23 ± 0.5) °C, measured immediately after powering on. This test was performed during the assembly phase of the 10 V prototype.

Table 3. Stability of the 10 V of an MRS measured at INRIM in the thermal chamber.

| Date | Relative deviation from nominal value ($\times 10^{-6}$) |
|---|---|
| 08/07 | 5.02 |
| 20/07 | 5.11 |
| 08/08 | 5.05 |
| 20/08 | 5.09 |
| 08/09 | 5.12 |
| 20/09 | 5.16 |

Table 4. INRIM measurements of a 10 V standard without thermal control after switching on.

| Time (h) | Measurements without thermal control from power on. Relative deviation from initial value ($\times 10^{-6}$) | Time (h) | Measurements without thermal control from power on. Relative deviation from initial value ($\times 10^{-6}$) |
|---|---|---|---|
| 1 | 0.03 | 8 | 0.16 |
| 2 | 0.09 | 9 | 0.15 |
| 3 | 0.11 | 10 | 0.18 |
| 4 | 0.14 | 11 | 0.20 |
| 5 | 0.17 | 12 | 0.18 |
| 6 | 0.15 | 13 | 0.17 |
| 7 | 0.18 | 14 | 0.16 |

The data in Table 4 show that the drift of the 10 V standard without thermal control is significantly higher during the first seven hours after power-up, at approximately $2.2 \times 10^{-8}$ per hour, and decreases to about $1.4 \times 10^{-9}$ per hour between the 8th and 14th hour. Comparing these values with those in Table 3, it can be concluded that the thermal control of the MRS, maintained at ± 0.1 °C, combined with a stabilisation period of at least 12 hours prior to measurements, significantly improves the stability of this standard. This time period is also consistent with the stabilisation of DMMs/MFCs according to the manufacturers. Figures 2a–d show the long-term stability of the three standards of an MRS, beginning from its assembly. The measurements were performed at MI in a laboratory, thermally regulated at (23 ± 0.5) °C. The data show a significant drift during the first year and a half, probably due to the stabilisation process of the components. After this initial stabilisation period, the drift decreases and becomes more in line with the manufacturer's stability specifications.

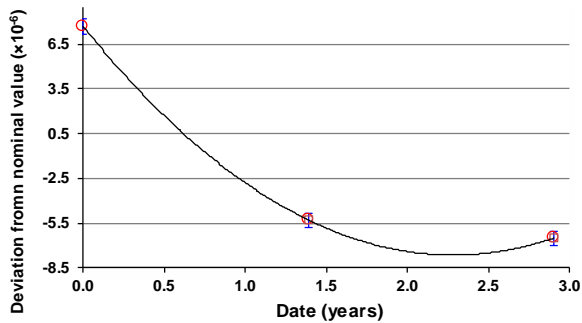 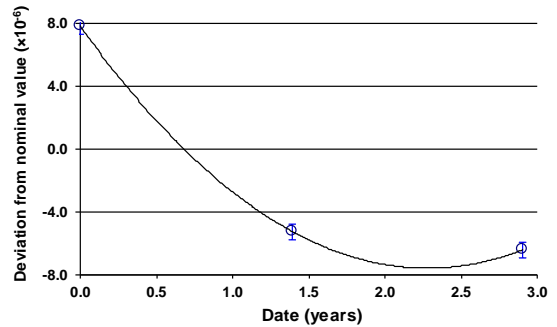

**Figure 2a.** Long-time stability of the 10 V of a MRS.   **Figure 2b.** Long-time stability of the -10 V of a MRS.

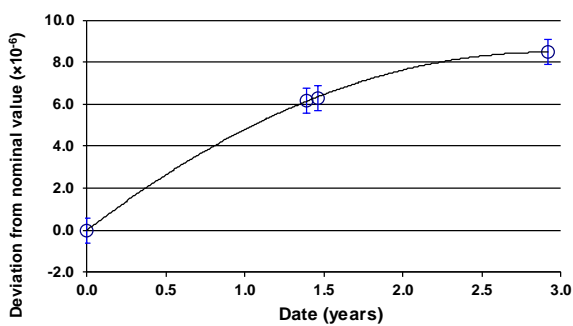 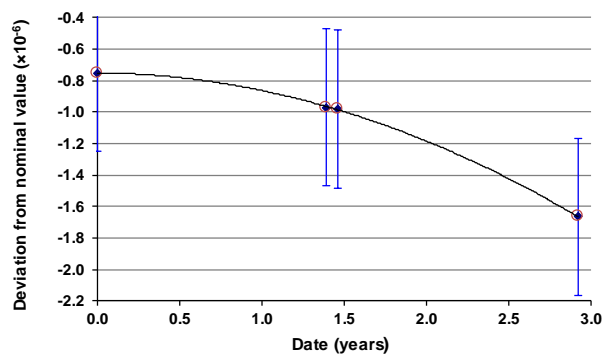

**Figure 2c**. Long-time stability of the 1 Ω of a MRS.  **Figure 2d.** Long-time stability of the 10 kΩ of a MRS.

## 4. Calibration of the MRS standards

The MRS standards must be calibrated with traceability to the SI through the DC voltage and DC resistance national standards according to the procedures and the quality system of the calibration laboratory. The 10 V standard can be calibrated either by comparison with a 10 V reference standard or through the Josephson effect [18] via a dedicated connection at room temperature, using a high-precision nanovoltmeter as a null detector. Similarly, the resistors can be calibrated with current comparator bridges by comparison with reference resistors traceable to the von Klitzing constant [19] or as a direct comparison to the quantum Hall effect using a cryogenic current comparator [20]. Table 5 reports the calibration results of the MRS standards obtained at INRIM during the period in which a MRS unit was available.

Table 5. Calibration results at INRIM of the standards of a MRS at 23 °C in laboratory environment.

| Nominal values | Calibration values | Measurement current mA | Relative expanded uncertainties ($k = 2$) ($\times 10^{-6}$) |
|---|---|---|---|
| 10V | 10.000051 V | | 0.5 |
| 1 Ω | 1.0000101 Ω | 100 | 0.5 |
| 10 kΩ | 9999.905 Ω | 0.1 | 0.5 |

## 5. Using the MRS as a calibration standard

Figure 3 shows a measurement setup for the artifact calibration of a MFC using the MRS standards. The connections and calibration steps are detailed in the user manuals of the instruments undergoing artifact calibration. For example, the Fluke 5720 MFC [21], after the artifact calibration and before saving the new constants, shows the changes of its internal references as ± part per million (ppm). Table 6 provides a short report of an artifact calibration carried out at INRIM using the MRS standards.

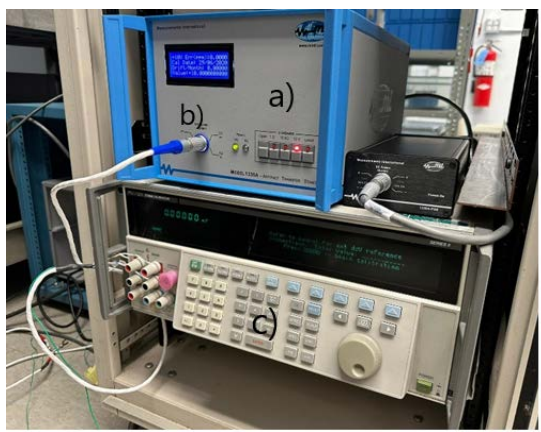

Figure 3. Photo of the measurement setup for the artifact calibration of a Fluke 5720 MFC (c) with the MRS (a). (b) is the LEMO connector. The connections to the LEMO connector are specifically designed to interface with the instruments in artifact calibration mode without requiring any modifications during the entire process. This design allows the calibration process to be fully automated, including zeroing.

Table 6. Short report of an artifact calibration of a Fluke 5720 MFC with the MRS standards at INRIM.

| Standard | Reference values | J. Fluke 5720 reference changes (ppm) | |
|---|---|---|---|
| 10 V MRS | 10.000051 V | 6.5 V | 0.3 |
|  |  | 13 V | -0.3 |
| 10 kΩ MRS | 9.999905 kΩ | 10 kΩ | 0.1 |
| 1 Ω MRS | 1.0000101 Ω | 1 Ω | -2.9 |
|  |  | 1.9 Ω | -3.7 |

## 6. Algorithm to provide the real-time value of the standards

In general, when a laboratory calibrates DMMs and MFCs using artifact calibration, it is recommended to perform this process as soon as possible after the calibration of the 10 V, 1 Ω, and 10 kΩ reference standards to minimise their drift. Since the MRS's integrated algorithm provides real-time values for its standards, there is no need to perform the artifact calibration within a specific time frame after the calibration of the reference standards. The calibration values of the standards are stored in the MRS's control EPROM. By means of a real-time clock (RTC) and the algorithm, the MRS displays both the calibration values of the standards and their updated values. These updated values are calculated through an extrapolation process based on prior calibrations. The current version of the algorithm is simple: the user enters the calibration value of each standard, the calibration dates, and the evaluated annual drifts. The algorithm then updates the values of the standards using a linear fit defined by equation (1).

$$X_d = X_{cal} + \left(d \frac{D}{365}\right) \quad (1)$$

where:
- $X_d$ is the value of a selected standard at the day $d$;
- $X_{cal}$ is the last calibration value of the standard;
- $D$ is the estimated annual drift of the standard in relative parts;
- $d$ the number of days after the last calibration.

## 7. Measurement uncertainties

When an MRS standard is used in the time period between two its calibrations, its associated in-use uncertainties must be considered[2] [22]. These uncertainties include contributions from calibration, drift, temperature variations, power, and transport effects. The implementation of the MRS algorithm significantly reduces the uncertainty component due to the drift. Table 7 shows the in-use uncertainty budgets for the three standards. The transport effect on the 10 V standard of an MRS, which is the most sensitive standard, was evaluated during an ILC among Canadian laboratories. It resulted less than $4.0 \times 10^{-8}$.

Table 7 In-use uncertainty budget according to [23] of the MRS standards (negl. indicates that the uncertainty contribution is negligible).

| Standard | 10 V | 1 Ω | 10 kΩ | Type and origin |
|---|---|---|---|---|
| Uncertainty component | Relative uncertainty ($k = 1$) | | | |
| Calibration | $2.5 \times 10^{-7}$ | $2.5 \times 10^{-7}$ | $2.5 \times 10^{-7}$ | Normal B INRIM calibration |
| Drift | $2.4 \times 10^{-9}$ | $7.9 \times 10^{-10}$ | $7.9 \times 10^{-10}$ | Rect. B MI spec. + algorithm |
| Temperature effect | $5.8 \times 10^{-8}$ | $5.8 \times 10^{-8}$ | $5.8 \times 10^{-8}$ | Rect. B MI spec.+ thermal reg. |
| Emf | $1.2 \times 10^{-8}$ | $1.2 \times 10^{-7}$ | negl. | Rect. B 20 μV typical value |
| Power effect | − | $1.4 \times 10^{-7}$ | $1.0 \times 10^{-7}$ | Rect. B MI spec. |
| Pressure effect | negl. [3] | negl. | negl | Rect. B [14] |
| Transport effect | $2.3 \times 10^{-8}$ | $2.3 \times 10^{-8}$ | $2.3 \times 10^{-8}$ | Rect. B from ILC |
| $u_{\text{in-use}}(x_i)$ ($k =1$) | $2.5 \times 10^{-7}$ | $3.1 \times 10^{-7}$ | $2.5 \times 10^{-7}$ | |
| $U_{\text{in-use}}(x_i)$ ($k =2$) | $0.5 \times 10^{-6}$ | $0.6 \times 10^{-6}$ | $0.5 \times 10^{-6}$ | |

Assuming[4] a relative expanded ($k =2$) calibration uncertainty of $0.5 \times 10^{-6}$, the expanded in-use uncertainties are $0.5 \times 10^{-6}$, $0.6 \times 10^{-6}$ and $0.5 \times 10^{-6}$ for the 10 V, 1 Ω and 10 kΩ standards, respectively. The uncertainty component due to the drift is negligible since the extrapolation function of the MRS values reduces this component to:

$$u_{drift} = \frac{\left(\frac{D}{365}\right)}{\sqrt{3}} \qquad (2)$$

assuming a rectangular distribution with amplitude $2D/365$ for the drift effect where $D$ represents the annual drift. The values provided by the manufacturer (Table 1) have been used for $D$ in the uncertainty budget of Table 7. Similar considerations apply to the uncertainty component related to the standards' temperature dependence, which is reduced by the thermo-regulation of the MRS. The in-use uncertainties of the MRS standards also apply when they are used as laboratory references during the interval between their calibrations.

---

[2] The concept of use uncertainty has been introduced since the calibration uncertainty of a standard could not be valid after some time after calibration. This is due to the standard drift and to the conditions in which the standard is used that could be different from those in which it was calibrated.

[3] According to [14], the pressure coefficient is 0.025 μΩ/Ω in the 700–1200 hPa range, so, since the resistors are in a box and in a pressure-monitored laboratory, the contribution due to pressure can be considered negligible.

[4] According to the INRIM Calibration measurement capabilities (CMC).

**Conclusions**

The MRS is a versatile and efficient static and portable metrological reference unit. It represents an effective solution for the calibration (adjustment) of high-end digital multimeters and multifunction calibrators using artifact calibration. From a metrological point of view, the MRS specifications meet the uncertainty requirements of manufacturers of high-end MFCs and DMMs. Economically and logistically, the MRS offers significant advantages, making it a cost-effective and practical choice. The MRS can also be used as a travelling standard for ILCs for DC resistance and DC voltage measurements. Future efforts will focus on improving the data extrapolation algorithm to further increase the accuracy and reliability of the instrument as the number of calibrations increases.